
%
%
%
%
%
\magnification=\magstep0
\tolerance=2500
\baselineskip=11.0pt
\hsize=13.5truecm
\vsize=20.925truecm
\hoffset=1.2truecm
\voffset=1.3truecm
\font\title=cmb10 scaled \magstep1
\font\mib=cmmib10 scaled \magstep1

\def\mb#1{\hbox{\mib #1}}
\def\tJ{$\mb{t}$\title{-}$\mb{J}$~}

\def\geqap{\,\raise 2pt \hbox{$>\kern-11pt \lower 5pt \hbox{$\sim$}$}\,}
\def\leqap{\,\raise 2pt \hbox{$<\kern-10pt \lower 5pt \hbox{$\sim$}$}\,}
\noindent
{\title Exact Diagonalization Study of Strongly Correlated Electron Models:}
\smallskip\par\noindent
{\title Hole pockets and shadow bands in the doped \tJ model}
\bigskip\noindent
{\sl Y. Ohta and R. Eder}
\bigskip\noindent
Department of Applied Physics, Nagoya University, Nagoya 464-01,
Japan
\bigskip\bigskip\noindent
{\bf Abstract.}
A detailed exact-diagonalization study is made for the doping
dependence of the single-particle spectral function $A({\bf k},\omega)$
and momentum distribution function $n({\bf k})$ of the two-dimensional
$t$$-$$J$ model as a representative model for doped Mott insulators.
The results for $A({\bf k},\omega)$ show unambiguously that the
rigid-band behavior is realized in the small-cluster $t$$-$$J$ model:
upon doping, the uppermost states of the quasiparticle band observed
at half filling simply cross the Fermi level and reappear as the
lowermost states of the inverse photoemission spectrum, while the
photoemission side of the band remains essentially unaffected.
We discuss problems in directly determining the Fermi surface from
$n({\bf k})$ and make a situation where they are largely avoided;
we then find clear signatures of a Fermi surface which takes the
form of small hole pockets.  The identical scaling with $t/J$ of the
quasiparticle weight $Z_h$ and difference in $n({\bf k})$ between
neighboring ${\bf k}$-points suggests the existence of such a Fermi
surface in the physical regime of parameters.
We construct spin-bag operators which describe the holes dressed
by the antiferromagnetic spin fluctuations and find that elementary
electronic excitations of the system can be described well in terms
of weakly-interacting spin-1/2 Fermionic quasiparticles corresponding
to the doped holes.
We make a comparison with other numerical calculations and recent
angle-resolved photoemission experiment and argue that, adopting this
rather conventional Fermi-liquid scenario with non-Luttinger Fermi
surface, one would explain many quasiparticle-relating properties of
doped cuprates in a very simple and natural way.  We also show that
the dynamical spin and charge excitations deviate from the
particle-hole excitations of this Fermi liquid: e.g., the dominant
low-energy spin excitation at momentum transfer $(\pi,\pi)$ reflects
excitations of the incoherent spin background and is identified
as a collective mode comparable to spin waves in the Heisenberg
antiferromagnet.

\bigskip\bigskip\noindent
{\bf 1.~~Introduction}
\bigskip
Physics of correlated electrons near a Mott-Hubbard metal-to-insulator
transition has been of great interest in recent years after the
discovery of high-temperature superconductors.  Here a well-known
particularly intriguing problem is the volume of the Fermi surface
(FS) for the slightly less than half-filled band.  Should one model
the doped insulator by a dilute gas of quasiparticles corresponding
to the doped holes [this would imply that the volume of the FS is
proportional to the hole concentration] or can the ground state still
be obtained by adiabatic continuation from the noninteracting one
with all electrons taking part in the formation of the FS so that its
volume is identical to that of free electrons?  Based on numerical
studies of the two-dimensional (2D) $t$$-$$J$ model, we present
evidence that the first picture is a correct one [1--3]: the
calculated photoemission spectrum can be interpreted very well in
the rigid-band approximation (RBA), the low-energy electronic
excitations can be described as holes dressed by antiferromagnetic
spin fluctuations as expected in the string or spin bag picture,
and the FS with the form of small hole pockets can be seen clearly
in the momentum distribution function if obvious problems are
circumvented.

This non-Luttinger `small' FS picture, shown [4,5] to be consistent
with majority of transport properties in the normal state of
high-temperature superconductors, was discarded some time ago because
of the apparent contradiction to the angle-resolved photoemission
(ARPES) experimet where the existence of a `large FS' consistent with
the LDA band-structure calculations was claimed.  Most recently,
however, this picture has been put under reconsideration; results of
newly reported ARPES (and other) experiment [6,7] can be interpreted
as indications of the small FS picture, as we will discuss in \S6.
Some doubts have also been casted to numerical studies of the $t$$-$$J$
and Hubbard models [8--11]: e.g., a recent quantum Monte Carlo
calculation suggests the existence of such a hole-pocket--like FS in
the 2D Hubbard model [11].  Here we report results of our
exact-diagonalization studies of the $t$$-$$J$ model.

The $t$$-$$J$ model is defined by the Hamiltonian
$$
H=-t\sum_{<ij>\sigma}
(\hat c^\dagger_{i\sigma}\hat c_{j\sigma}+{\rm H.c.})
+J\sum_{<ij>}({\bf S}_i\cdot {\bf S}_j - {1\over 4}n_i n_j)
\eqno(1)
$$
where the operators $\hat c_{i\sigma}$ are expressed in terms of
ordinary Fermion operators as $c_{i\sigma}(1-n_{i\bar\sigma})$,
$n_i$$=$$n_{i\uparrow}$$+$$n_{i\downarrow}$ are the electron number
operators, and ${\bf S}_i$ are the electronic spin operators.
The summation is taken over all the nearest-neighbor pairs
$<$$ij$$>$ on the 2D square lattice.  The photoemission (PES) and
inverse photoemission (IPES) spectra are defined respectively as
$$\eqalign{
&A_n^-({\bf k},\omega)=\sum_\nu
|\langle\Psi_\nu^{(n+1)}|c_{\bf k\sigma}|\Psi_0^{(n)}\rangle|^2
\delta(\omega-(E_\nu^{(n+1)}-E_0^{(n)})) \cr
&A_n^+({\bf k},\omega)=\sum_\nu
|\langle\Psi_\nu^{(n-1)}|c_{\bf k\sigma}^\dagger|\Psi_0^{(n)}\rangle|^2
\delta(\omega-(E_\nu^{(n-1)}-E_0^{(n)}))
}
\eqno(2)
$$
with the single-particle spectral function
$A({\bf k},\omega)$$=$$A_n^-({\bf k},-\omega)$$+$$A_n^+({\bf k},\omega)$.
Here $|\Psi_\nu^{(n)}\rangle$ ($E_\nu^{(n)}$) is the $\nu$-th eigenstate
(eigenenergy) with $n$ holes ($\nu$$=$$0$ implies the ground state).

\bigskip\bigskip\noindent
{\bf 2.~~Problems in Identifying the Fermi Surface}
\bigskip
Various authors have computed the momentum distribution function
$n({\bf k})$$=$$\langle\hat c^\dagger_{{\bf k}\sigma}
\hat c_{{\bf k}\sigma}\rangle$
and single-particle spectral function $A({\bf k},\omega)$ for the
two-hole ground state of small clusters of this model (corresponding
to a nominal doping of $\sim$10\%), and found that $n({\bf k})$ is
maximum at ${\bf k}$$=$$(0,0)$ and minimum at
${\bf k}$$=$$(\pi,\pi)$ and roughly consistent with a free-electron
picture and a quasiparticle-like band structure can be assigned in
$A({\bf k},\omega)$ in the neighborhood of the Fermi energy which
resembles that for noninteracting particles.  It has become customary
[12] to cite this as evidence that already at such fairly low doping
levels the $t$$-$$J$ model has a free-electron--like `large' FS.
However, it is straightforward to see that this shape of $n({\bf k})$
is simply the consequence of elementary sum rules and has no significance
for the actual topology of the FS [10]: The expectation value of the
kinetic-energy term of Eq.~(1) may be expressed as
$$
\langle H_t \rangle = \sum_{{\bf k}\in {\rm AFBZ}}
\varepsilon ({\bf k})\delta n({\bf k})
\eqno(3)
$$
with $\delta n({\bf k})$$=$$n({\bf k})$$-$$n({\bf k}+{\bf Q})$,
where $\varepsilon({\bf k})$ is the free-particle energy,
${\bf Q}$$=$$(\pi,\pi)$, and the summation is restricted to inside
of the antiferromagnetic Brillouin zone (AFBZ).  Since
$\varepsilon({\bf k})$$<$$0$ inside the AFBZ, the average of
$\delta n({\bf k})$ must be positive there to ensure
$\langle H_t \rangle$$<$$0$.  Also, denoting by $M_0^-$ ($M_0^+$)
the 0th moment of the PES (IPES) spectrum, we have
$M_0^-$$=$$n({\bf k})$ and $M_0^+$$=$${\rm const}$$-$$n({\bf k})$.
The mere requirement $\langle H_t \rangle$$<$$0$ thus enforces that
(i) $n({\bf k})$ be larger inside the AFBZ than outside and (ii)
IPES weight predominantly appear outside the AFBZ and PES weight
be concentrated within the AFBZ.  Thus, such `nearest-neighbor hopping'
band and `free-electron--like' FS mainly reflect generic properties
of any Hamiltonian where the kinetic energy is a nearest-neighbor
hopping term.  It is also easy to construct a counterexample of a
system, where the dispersion of lowest-energy sharp peaks can fit
to the nearest-neighbor hopping band, while it is guaranteed that a
free-electron--like FS is rigorously excluded due to a build-in
broken symmetry [10].

We must therefore consider how the more subtle questions, such as
the existence and shape of the FS and the band structure near
the Fermi energy, can be answered in our numerical method.
Besides the above elementary sum rules, the key problems which we
have to bear in mind are as follows:

A:~For strongly correlated systems, the `quasiparticle peak' near
the chemical potential carries only a small fraction $Z_h$ of the
total weight of the single-particle spectral function; in other words,
the small `Fermi-edge discontinuity' in $n({\bf k})$, which has to
be equal to $Z_h$, is superimposed over a substantial background
steming from the integration of the incoherent continua.  Then, when
only $n({\bf k})$ is considered, a systematic change in the weight of
the background, which is unrelated to low-energy physics, may mimic
an FS.

B:~A variety of diagonalization studies have shown that two holes
in the $t$$-$$J$ model form a bound state with a binding energy
($E_B$$\simeq$$0.8J$$-$$J$) that is a sizeable fraction of the
single-hole bandwidth ($W$$\simeq$$2J$).  The ground state with
two `holelike quasiparticles' should thus be modeled by a state of
the type
$$
|\Psi_0\rangle=\sum_{\bf k}\Delta({\bf k})
a_{{\bf k}\uparrow}^\dagger a_{-{\bf k}\downarrow}^\dagger
|{\rm vac}\rangle
\eqno(4)
$$
where $a_{{\bf k}\sigma}^\dagger$ is the quasiparticle creation
operator in the undoped vacuum state $|{\rm vac}\rangle$.  The
wave function $\Delta({\bf k})$ may differ appreciably from zero
for all quasiparticle states within $\sim$$E_B$ above the ground
state: no signatures of the FS can be seen unless $\Delta({\bf k})$
is well localized in {\bf k}-space, i.e.,
$\Delta({\bf k})$$\sim$$\delta_{{\bf k},{\bf k}_0}$.
In this situation, the quasiparticle peak at any momentum splits
into two peaks, one in the IPES spectrum with intensity proportional
to the quasiparticle occupation $\tilde n({\bf k})$ and the other
in the PES spectrum with the intensity proportional to
$1$$-$$\tilde n({\bf k})$.  Adding up these weight, one should then
ideally obtain the weight of the `unsplit' quasiparticle peak.
Moreover, the $d_{x^2-y^2}$ symmetry of the two-hole ground state
(in, e.g., 16 site- and 18-site clusters) implies that $\Delta({\bf k})$
must have a node along $(1,1)$ direction and hence
$\tilde n({\bf k})$$=$$0$ for these momenta; the peak splitting should
thus occur predominantly near $(\pi,0)$.

C:~There is a (near) degeneracy of the quasiparticle despersion
along the surface of the AFBZ, which is unfavorable in identifying
the location of the quasiparticles in ${\bf k}$-space: a `dilution'
occurs because few holes are distributed over many ${\bf k}$-points
with almost the same energy.  A marginal next-nearest-neighbor hopping
term $t'$ may be included in the Hamiltonian to lift this degeneracy
(see \S5):  The negative sign of $t'$, in particular,  favors hole
occupancy of ${\bf k}$$=$$(\pi,0)$ which has a low multiplicity
favorable for avoiding the additional dilution.  Moreover, in the
16-site cluster, this term breaks the spurious symmetry due to the
mapping of a $2^4$ hypercube and selects a unique two-hole ground
state with momentum $(0,0)$.

\bigskip\noindent
{\bf 3.~~Rigid-Band Picture}
\bigskip
The RBA states that, upon doping, the chemical potential shifts across
the quasiparticle band at half filling, while no change occurs in the
quasiparticle band structure; the lowermost peaks of the IPES spectrum
for the doped case are identical as the uppermost states in the PES
spectrum for the undoped case.  Let us see whether this is realized in
the spectra obtained by the exact diagonalization.

Figure 1 compares $A({\bf k},\omega)$ in the half-filled and
two-hole ground states for all allowed momenta in the 16- and 18-site
clusters.  We find that along the $(1,1)$ direction, where the pair
wave function $\Delta({\bf k})$ vanishes, there is a striking similarity
between the PES spectra near the Fermi energy $E_{\rm F}$ for the doped
and undoped cases: a band of peaks with practically identical dispersion
and weight can be clearly identified in both groups of spectra
[a possible exception is $(2\pi/3,2\pi/3)$: the single-hole ground states
of this momenta and $(\pi,\pi)$ have total spin $S$$=3/2$].
Away from $(1,1)$, doping leads to a shift of weight from the PES
band to IPES peaks immediately above $E_{\rm F}$; as expected from the
symmetry of $\Delta({\bf k})$, this shift is most pronounced near $(\pi,0)$
(see B of \S2).  The results for the 20-site cluster also indicate
the same similarity between the low-energy parts of the PES spectra
for the doped and undoped cases: the dominant low-energy PES peaks
near $E_{\rm F}$ remain either unaffected or (partially) cross the
chemical potential to reappear as lowest-energy IPES peaks.  At
energies remote from $E_{\rm F}$ (and thus unrelated to any low-energy
physics), there is a reshuffling of incoherent spectral weight [pronounced
addition at $(0,0)$ and depletion at $(\pi,\pi)$], which causes
the formation of a `large FS'--like $n({\bf k})$.

\baselineskip=10.0pt
\midinsert \vskip9.0truecm
\hsize=12.5truecm \leftskip=1.0truecm
\noindent
Fig.~1.~~Comparison of the photoemission spectra
at half filling and in the ground state with two holes:
$A_0^-({\bf k},-\omega)$ (soild line),
$A_2^-({\bf k},-\omega)$ (dotted line), and
$A_2^+({\bf k},\omega)$ (dot-dashed line)
for the 16- and 18-site clusters with $J/t$$=$$0.4$ are shown.
The Fermi energy is marked by the vertical line,
and $\delta$ functions have been replaced by Lorentzian
of width $0.1t$.
\endinsert
\baselineskip=11.0pt

A quantitative check of the validity of RBA is to see whether the
lowest-energy IPES peaks in the two-hole ground state exactly observe
the single-hole quasiparticle band observed in PES at half filling.
We choose $E_0^{(2)}$ as the `ground-state' energy in the PES
spectrum at half-filling and do not invert the sign of $\omega$:
we can thus make a direct comparison of the positions of peaks in
this spectrum and in the IPES spectrum for the two-hole ground state
because both spectra involve the single-hole subspace in their final
states.  The calculated results in the 16-, 18-, and 20-site
clusters confirm that the final states for the lowest IPES peaks
at all momenta off the $(1,1)$ direction indeed belong to the
single-hole band observed in PES spectrum at half filling within
an accuracy of $10^{-10}t$ [there is no low-energy IPES peak along
the $(1,1)$ direction because $\tilde n({\bf k})$$=$$0$ (see B in
\S2)].  It is only at higher energies ($\sim$$J$ above the
quasiparticle states) that there is a significant difference in
the spectra.  Note that this result is in clear contradiction to
the `large FS' scenario: this would necessitate the assumption
that the uppermost states of the next-nearest-neighbor hopping
band at half filling simultaneously belong to a topologically
different nearest-neighbor hopping band obtained under a `full-scale
transition' upon doping of two holes.

Another quantitative check of the RBA is to compare the spectral
weight of the peaks near $E_{\rm F}$.  Figure 2 (a) compares the
weight of the peak at $(\pi/2,\pi/2)$ in $A_0^-({\bf k},\omega)$
[which has the same weight at $(\pi,0)$] and $A_2^-({\bf k},\omega)$
as well as the sum of the weights of the lowest peak in
$A_2^+({\bf k},\omega)$ and highest peak in $A_2^-({\bf k},-\omega)$
at $(\pi,0)$ [which are in the `split-peak' situation].  The RBA
predicts all three quantities to be equal, and as seen in Fig.~2 (a),
they are indeed agree remarkably well over a wide range of $t/J$.

\baselineskip=10.0pt
\midinsert \vskip4.7truecm
\hsize=12.5truecm \leftskip=1.0truecm
\noindent\hskip2.8truecm (a) \hskip5.5truecm (b)
\medskip
\noindent
Fig.~2.~~(a) Comparison of the $t/J$ dependence of the PES pole
strength at $(\pi/2,\pi/2)$ at half filling (squares) and in the
two-hole ground state (triangle) and the added weights of the
lowest IPES and highest PES peak at $(\pi,0)$ in the two-hole
ground state (circles).
(b) Quasiparticle band structure in the neighborhood of $E_{\rm F}$
for the 16- and 18-site cluster with $J/t$$=$$0.4$.
Uptriangles (squares) give the position of the highest PES peak,
and down triangles (circles) give the position of the lowest IPES
peak for the 16-site (18-site) cluster with two holes.  The positions
of the highest PES peak at half filling are also given by dots.
The various groups of spectra have been shifted so that the energies
of the respective PES peak at $(0,0)$ coincide [shift between doped
16- and 18-site clusters is $0.275t$]. A box indicates the peaks
split due to hole binding.
\endinsert
\baselineskip=11.0pt

The quasiparticle band structure near $E_{\rm F}$ is summarized in
Fig.~2 (b):  The dispersion in the two-hole states agree very well
with that of the half-filled band, as the RBA predicts.  Also,
comparing Fig.~1 with 2 (b), we note an obvious correlation
between the quasiparticle peak intensity and the distance from
$E_{\rm F}$, as one would expect for a Fermi liquid.  Due to the
interaction between the holes, there is no FS but rather a zone of
partially occupied momenta where the quasiparticle peaks are split
between PES and IPES.  However, the obvious validity of the RBA
suggests that, if the FS exists at all, it takes the form of small
hole pockets, although, due to the near degeneracy of the states on
the surface of the AFBZ, the precise location of the pockets are
decided by the interaction between holes.   Along the $(1,1)$ direction,
another band of many-body states with intrinsic $d_{x^2-y^2}$ symmetry
is also identified (see Ref.~[1] for details): this band has almost no
dispersion and remains $\sim$$2J$ above $E_{\rm F}$, and thus is
unrelated to any low-energy physics.

We construct a `spin-bag' operator to describe the low-energy
states thus found near $E_{\rm F}$, in terms of elementary
excitations, i.e., weakly or moderately interacting quasiparticles,
whereby the spectral function becomes an almost free-particle form
with the incoherent continua being removed and the peaks near
$E_{\rm F}$ being enhanced. Because a `single-hole problem' can be
described well by the string picture where the hole is dressed by
antiferromagnetic spin fluctuation, we make the ansatz for the
operator:
$$
\tilde c_{{\bf k}\uparrow}=\sum_{\lambda=0}^{\lambda_{\rm max}}
\alpha_\lambda ({\bf k}) A_\lambda ({\bf k})
\eqno(5)
$$
where $A_\lambda ({\bf k})$ is the Fourier transform of
$A_\lambda ({\bf R}_j)$ which creates all strings of length $\lambda$
around site $j$ when acting on the N\'eel state (see Ref.~[2] for
details).  From the spectral functions for $\tilde c_{{\bf k}\sigma}$
with coefficients $\alpha_\lambda$ determined variationally, we find
the following:  (i) The spin-bag operator optimized at half filling
works well even for the two-hole case: this suggests the continuity
in the development of low-energy states upon doping to the single-hole
problem in an antiferromagnet.  (ii) The Pauli principle works for the
spin-bag operator: in the spectrum of adding a bag with up-spin
from the one-hole ground state [containing 8 up-spin and 7 down-spin
electrons in the 16-site cluster] at momentum $(\pi/2,\pi/2)$, there
is a large elimination of the incoherent continua and enhancement of
the peaks at $E_{\rm F}$, whereas in the spectrum of adding a bag
with down-spin, the hole pocket is clearly seen at $(\pi/2,\pi/2)$.
(iii) The degree of broadening of the spin-bag peaks is reminiscent of
a Fermi liquid: there are sharp peaks close to $E_{\rm F}$ and diffuse
peaks at lower energies.  (iv) We find an approximate adjoint (or
spin-bag annihilation) operator which works well to simplify the IPES
spectrum.  Thus, all in all the doped cluster behaves very much like
a system of weakly interacting `effective Fermions' corresponding to
the doped holes.

\bigskip\noindent
{\bf 4.~~Hole Pockets}
\bigskip
We now present evidence that the FS as deduced from the momentum
distribution function takes the form of small hole pockets.  We
first note that the magnitude of the Fermi-edge discontinuity in
$n({\bf k})$ has to be equal to the weight $Z_h$ of the quasiparticle
peak in $A({\bf k},\omega)$.  Then, to identify the FS, we can take
advantage of the fact that $Z_h$ has a pronounced (and characteristic)
dependence on $t/J$ [13]: i.e., a potential FS discontinuity must
have the same characteristic dependence on $t/J$.

Figure 3 (a) shows $n({\bf k})$ calculated for the single-hole
ground state with momentum ${\bf k}_0$$=$$(\pi/2,\pi/2)$.  We
find the ${\bf k}$-dependence roughly consistent with free
electrons, i.e., larger near $(0,0)$ and smaller near $(\pi,\pi)$,
which ensures the negative kinetic energy (see \S2), and also
we note a dip at ${\bf k}_0$ for the minority-spin $n({\bf k})$.
The question arises which of these features should be associated
with the FS: do we have a `large FS' already for a single hole
or is there a `hole pocket' at ${\bf k}_0$?  Figure 3 (b) shows
a comparison between the `depth' of the dip at ${\bf k}_0$
($\Delta_{\rm dip}$) [estimated by forming the difference in $n({\bf k})$
with a symmetry equivalent ${\bf k}$ point] and the weight of the
quasiparticle peak [obtained from the single-particle spectral function
for momentum transfer ${\bf k}_0$ at half filling] for various values
of $t/J$.  Obviously, $\Delta_{\rm dip}$$=$$Z_h$ over the entire range
of $t/J$, so that the dip clearly originates from the Fermi level
crossing of the quasiparticle band, i.e., we have a hole pocket at
${\bf k}_0$.  On the other hand, differences $\Delta n({\bf k})$ across
the `large FS' always show the opposite behavior under a variation of
$t/J$ as $Z_h$, indicating that these drops in $n({\bf k})$ are
unrelated to any FS crossing; it seems reasonable to associate this
structure in $n({\bf k})$ with the well-known backflow for interacting
Fermi systems [14].

\baselineskip=10.0pt
\midinsert \vskip7.0truecm
\hsize=12.5truecm \leftskip=1.0truecm
\noindent\hskip2.6truecm (a) \hskip5.3truecm (b)
\medskip
\noindent
Fig.~3.~~(a) Momentum distribution for the single-hole ground state
with $S_z$$=$$1/2$ (i.e., with a `down' hole) for the 16-site cluster
with $t/J$$=$$4$ (upper panel) and $t/J$$=$$1$ (lower panel).  The
upper (lower) values in the lists refer to the majority (minority)
spin, and the ground-state momentum ${\bf k}_0$ is marked by a black
box and ${\bf k}_0$$+$$(\pi,\pi)$ by a dotted box.
(b) Comparison of the $t/J$ dependence of $Z_h$ at half filling
(dark squares) and various differences $\Delta n({\bf k})$ in the
single-hole ground states with $S_z$$=$$1/2$.  Shown is the `depth'
of the pockets (light circles) and differences across the `large FS'
(up and down triangles).
\endinsert
\baselineskip=11.0pt

Let us proceed to the two-hole case.  A free-electron--like
variation of $n({\bf k})$ is clearly seen in, e.g., the 20-site cluster
and thus one would be tempted to assign this to the existence of
a `Luttinger FS' by adopting a criterion like $n({\bf k})$$>$$1/2$
[12].  However, such a Luttinger FS is ruled out by the same arguments
as for a single hole: Fig.~4 (b) compares between the $t/J$ dependence
of $\Delta n({\bf k})$ across the respective Luttinger FS and that of
the quasiparticle weight $Z_h$ in the spectral function for the two-hole
ground state.  $Z_h$ decreases sharply, while $\Delta n({\bf k})$
increase monotonically with $t/J$.  Thus, the drop in $n({\bf k})$
upon crossing the `large FS' is obviously unrelated to any true
Fermi-level crossing.  We note [3] that, if we assume that the backflow
contribution for the two holes is simply additive of that for the
single hole, the magnitude and $t/J$ scaling of $\Delta n({\bf k})$
can be explained very well.

\baselineskip=10.0pt
\midinsert \vskip6.3truecm
\hsize=12.5truecm \leftskip=1.0truecm
\noindent\hskip0.7truecm (a) \hskip3.4truecm (b) \hskip4.2truecm (c)
\medskip
\noindent
Fig.~4.~~(a) Allowed momenta and `Luttinger FS' for various clusters.
The `Fermi momenta' are denoted by ${\bf k}_{\rm F}$.
(b) Comparison of the $t/J$ dependence of $Z_h$ (dark squares)
[obtained from the photoemission spectra at the ${\bf k}_{\rm F}$
indicated in (a)] and the difference $\Delta n({\bf k})$ (light
squares and circles) between pairs of momenta connected by dashed
lines in (a).
(c) Comparison of the scaling of $Z_h$ (dark squares) and selected
differences $\Delta n({\bf k})$ (light squares) with $t/J$: shown are
$1.5\cdot [n(\pi/2,\pi/2)$$-$$n(\pi,0)]$ for 16-site,
$2.8\cdot [n(\pi/3,\pi/3)$$-$$n(2\pi/3,0)]$ for 18-site, and
$3.6\cdot [n(\pi/5,3\pi/5)$$-$$n(\pi,0)]$ for 20-site clusters.
\endinsert
\baselineskip=11.0pt

Thus, we expect that the hole pocket is superimposed over the smooth
backflow contribution as in the single-hole case, only with the
additional complication that the pockets are now `washed out' due to
the interaction between holes.  Hence, $n({\bf k})$ may be written as
$$
n({\bf k})=n_{\rm back}({\bf k})-Z_h\cdot|\Delta({\bf k})|^2
\eqno(6)
$$
with the pair wave function $\Delta({\bf k})$ introduced in Eq.~(4).
Because the calculated results suggests that to a good approximation
the backflow contribution $n_{\rm back}({\bf k})$ is a function of
$|k_x|$$+$$|k_y|$ only (see, e.g., Fig.~3 (a)), this contribution
can be eliminated by forming the difference of two momenta with
(almost) equal $|k_x|$$+$$|k_y|$.  Next, if we choose one of these
momenta along (or near) the $(1,1)$ direction [where
$\Delta({\bf k})$ vanishes (or is small)] and the other at (or near)
$(\pi,0)$, we should obtain
$\Delta n({\bf k})$$=$$Z_h \cdot |\Delta(\pi,0)|^2$,
so that, in contrast to the `large FS' differences indicated in
Fig.~4 (b), this difference should scale with $Z_h$.  To check
this prediction, the $t/J$ dependence of various such differences
is shown in Fig.~4 (c): obviously, to an excellent approximation,
they are proportional to $Z_h$ over a wide range of $t/J$.  The
scaling of $n({\bf k})$ with $t/J$ is thus completely consistent
with the assumptions that (i) there are washed-out hole pockets at
$(\pi,0)$ and (ii) these are superimposed over the smooth backflow
contribution which is the sum of the backflows for the two individual
holes.

Now the last question would be whether one can really make the true
FS visible in the structure of $n({\bf k})$. The answer is simply to
circumvent the problems listed in A, B, and C of \S2:  For A,
since diagonalization studies have shown that $Z_h$$\simeq$$1$ for
the parameter region $J$$\geqap$$t$, we choose $J/t$$=$$1$ and $2$.
For B, we introduce a density repulsion term
$H_V$$=$$V\sum_{<ij>}n_in_j$ and adjust $V$-value so as to cancel the
intrinsic attractive interaction between holes and obtain homogeneous
distribution of holes.  And for C, we introduce a next-nearest-neighbor
hopping term with a fixed strength $t'$$=$$-0.1t$ to lift the (near)
degeneracy of the momenta along the surface of the AFBZ.  Thus, we
have the `ideal' situation for observing the FS: large Fermi-edge
discontinuity, weak interaction, and a unique single-hole ground state.
The results are shown in Fig.~5:  In the spectral function, $E_{\rm F}$
is located near the top but within a group of pronounced peaks which are
well separated from another such group in the IPES spectrum.  There are
pronounced peaks both immediately above and below $E_{\rm F}$ which
comprise the bulk of spectral weight for each momentum, indicating
well-defined quasiparticle peaks.  Correspondingly, $n({\bf k})$ exhibits
a sharp variation: i.e., there are hole pockets at $(\pi,0)$ and $(0,\pi)$.
They are superimposed over the familiar backflow contribution, which
again has the generic free-electron--like form to ensure negative
kinetic energy.  Figure 5 also gives the values of the quasiparticle
weight at the `Fermi momenta': we note that the depth of the pockets
approximately equals $Z_h$ and both quantities consistently decrease
with decreasing $J/t$.  The location of the pockets at $(\pi,0)$ rather
than at $(\pi/2,\pi/2)$ might be somehow surprising.  This can be traced
back to the point-group symmetry of the two-hole ground state: when the
symmetry of the ground state at half filling is $A_1$ (or $s$), that
of the two-hole ground state is $B_1$ (or $d_{x^2-y^2}$) and vice
versa.  Thus, addition of two holes is equivalent to adding an object
with $d_{x^2-y^2}$ symmetry.  This implies that the pair wave function
$\Delta({\bf k})$ in Eq.~(4) should have this symmetry as well, and
in turn implies that $\Delta({\bf k})$$=$$0$ for all ${\bf k}$'s along
the (1,1) direction, so that the hole occupation of $(\pi,0)$ is
favored.

\baselineskip=10.0pt
\midinsert \vskip7.4truecm
\hsize=12.5truecm \leftskip=1.0truecm
\noindent\hskip2.8truecm (a) \hskip5.1truecm (b)
\medskip
\noindent
Fig.~5.~~(a) Single-particle spectral function for the
$t$$-$$t'$$-$$J$$-$$V$ model with two holes: $J/t$$=$$2$ and
$V/t$$=$$2.5$ ($2.4$) are used for the 20-site (16-site) cluster.
The upper four rows refer to the 20-site cluster, and the lower
three rows to the 16-site cluster.  The frequency region
$\omega$$<0$ ($\omega$$>$$0$) corresponds to the PES (IPES)
spectrum.  Delta functions have been replaced by Lorentzian of
width $0.05t$.
(b) Momentum distribution in the two-hole ground state of the
$t$$-$$t'$$-$$J$$-$$V$ model.
In the upper panel, $J/t$$=$$2$ and $V/t$$=$$2.5$ ($2.4$) are
used for the 20-site (16-site) cluster, and
in the lower panel, $J/t$$=$$1$ and $V/t$$=$$3.0$ ($2.0$) are
used for the 20-site (16-site) cluster.
For the `Fermi momenta', the quasiparticle weight $Z_h$ is given
in brackets.
\endinsert
\baselineskip=11.0pt

A possible explanation for the hole-pocket FS could be
spin-density-wave--type broken symmetry: although the ground states
under consideration are spin singlets, this might be realized if the
fluctuations of the staggered magnetization $M_s$ were slow as compared
to the hole motion, so that the holes move under the influence of an
`adiabatically varying' staggered field.  A possible criterion for this
situation would be $\tau_{\rm tr}\cdot\omega_{\rm AF}$$\ll 2$$\pi$,
where $\tau_{\rm tr}$ is the time it takes for a hole to transverse
the cluster and $\omega_{\rm AF}$ is the frequency of fluctuations of
$M_s$.  We estimate (for $J/t$$=$$2$) the group velocity of the holes
from the dispersion of the `quasiparticle peak' in the PES spectrum
[i.e., from the energies indicated by arrows in Fig.~5 (a) for the
20-site cluster and the peaks at $(\pi/2,0)$ and $(\pi/2,\pi/2)$ for
the 16-site cluster] and find $\tau_{\rm tr}$$\simeq$$2\pi/0.5t$
(or $\simeq$$2\pi/0.2t$) for the 20-site (or 16-site) cluster.
Typical frequencies for fluctuations of $M_s$ can be obtained from
its correlation function, which, up to a constant, equals the dynamical
spin susceptibility for momentum transfer $(\pi,\pi)$; a rigorous
lower bound on $\omega_{\rm AF}$ thus can be obtained by subtracting
the ground state energy from the energy of the lowest state with total
momentum $(\pi,\pi)$ with the same point-group symmetry as the ground
state.  This gives $\omega_{\rm AF}$$>$$0.9t$ (or $1.2t$) for the 20-site
(or 16-site) cluster, i.e., $\tau_{\rm tr}\cdot\omega_{\rm AF}$$>$$2\pi$.
`Almost static' N\'eel order can thus be ruled out as origin of the
small FS, even for this fairly large value of $J/t$.

As an additional check, we have introduced exchange terms $J'$
between second- and third-nearest neighbors to reduce the spin
correlations and again optimized the density repulsion term to
enable `free' hole motion.  For this (highly artificial) model,
we calculate the momentum distribution, density correlation function
$g({\bf R})$ for holes, and spin correlation function $S({\bf R})$.
We find [3] that the density correlation function is homogeneous
(i.e., no charge ordering), the spin correlations decay rapidly
(i.e., no long-range AF or spiral ordering), but still there are
unambiguous hole pockets in $n({\bf k})$.  The only possible conclusion
is that it is solely the large $Z_h$ that makes the pockets visible in
the large $J/t$ region, and not the onset of any kind of ordering.

While the hole pockets can be made clearly visible for large $J$,
the situation is more involved for $t$$>$$J$.  In this parameter
region, the small overlap between `quasiparticle' and `bare hole'
(as manifested by the small $Z_h$) makes the $V$-term (which couples
only to the bare hole) increasingly inefficient in enforcing a
noninteracting state: the two holes remain bound on second-nearest
neighbors up to fairly large values of $V$.  However, because the
scaling of $\Delta n({\bf k})$ with $t/J$ works very well over a
wide range of $t/J$ as we have shown above and also because the
overlap of the ground-state wave function with that at $J/t$$=$$1$
indicates no significant drop over $J/t$$=$$2$$-$$0.2$, it seems
reasonable to accept the continuity of the presence of hole pockets
to the physically realistic (smaller $J/t$-parameter) regime.

\bigskip\noindent
{\bf 5.~~Comparison with Other Numerical Studies and Experiment}
\bigskip
While we have shown the validity of RBA, the presence of
hole-pocket FS, and the spin-bag description for the quasiparticles,
their evidence is based solely on small-cluster studies.  A comparison
with other numerical calculations and experiments on cuprate (and other)
materials is therefore necessary.  As far as numerical studies on small
clusters are concerned, hole pockets and/or rigid-band behavior upon
doping have been continually suggested:  Poilblanc and Dagotto [15]
studied the PES spectrum for single-hole states in the $t$$-$$J$ model
and concluded that the two-hole ground state in the 16-site cluster
shows hole pockets at $(\pi,0)$, in agreement with the present results.
Castillo and Balseiro [16] computed the Hall constant and found its sign
near half filling to be consistent with a hole-like FS, i.e., with hole
pockets.  Gooding {\it et al.}~[17] studied the doping dependence of
the spin correlation function in clusters with special geometry and
also found indications of rigid-band behavior.  The situation is
quite similar for the Hubbard model:  While the generic
free-electron--like shape of $n({\bf k})$ found in earlier Monte Carlo
studies was initially considered as evidence against hole pockets,
more careful and systematic analysis [11] showed that hole pockets are
in fact remarkably consistent with the numerical data, their
nonobservation in the earlier studies being simply the consequence of
thermal smearing.  It seems fair to say that the available numerical
results for small clusters of both Hubbard and $t$$-$$J$ models, when
interpreted with care, are all consistent with the RBA and/or
hole-pocket FS.

Let us next discuss experimental results on high-temperature
superconductors assuming that the hole pockets found in the cluster
studies persist in the real materials.  The volume of the FS associated
with bands for the CuO$_2$ plane presents a well-known puzzle:  Early
ARPES experiments show bands which disperse towards the Fermi energy
and vanish at points in ${\bf k}$-space which are roughly located on
the free-electron FS corresponding to electron density $1$$-$$\delta$,
where $\delta$ is the hole concentration.  Transport properties, on
the other hand, can be modeled well [4,5] by assuming an FS with a
volume $\sim$$\delta$.  In a Fermi liquid, these apparently contradicting
quantities actually fall into distinct classes:  PES spectra depend on
$Z_h$, transport properties do not.  Hence, if one wants to resolve the
descrepancy entirely within a Fermi-lquid--like picture, the simplest
way would be to assume a `small FS' and explain the PES results by a
systematic variation of $Z_h$ along the band which forms the FS, similar
to the `shadow band' picture [18].  A trivial argument for such strong
${\bf k}$-dependence of the quasiparticle weight is that a distribution
of PES weight in the Brillouin zone (and hence $n({\bf k})$) that
resembles the noninteracting FS always optimizes the expectation value
of the kinetic energy, and therefore it is favorable if those parts of
the band structure, which lie inside the free-electron FS, have large
spectral weight, and the parts outside have small weight.  Then, it is
noticed that, in a recent ARPES study by Aebi {\it et al.}~[6], spectral
structures which are very consistent with such a shadow-band scenario
have indeed been observed.  Moreover, another key feature of the
dispersion relation for a single hole, namely the extended flat region
near $(\pi,0)$ (see Fig.~2 (b)), has also been found as a universal
feature of high-temperature superconductors [19--21].  Thus, adopting
a rigid band/hole pocket scenario would explain many experiments in a
very simple and natural way, which is moreover remarkably consistent
with the existing numerical data as a whole.  A recent finding of
noncopper quasi-2D materials exhibiting similar anomalies to cuprates
[22] seems to suggest that features discussed above should be generic
and common to a class of doped Mott insulators.

\bigskip\noindent
{\bf 6.~~Conclusions}
\bigskip
We have performed a detailed study of the doping dependence of
the single-particle spectral function and momentum distribution
function up to the largest clusters that are numerically tractable.
The results show unambiguously that the rigid-band behavior is
realized in small clusters of the $t$$-$$J$ model: Near the
chemical potential, the main effect of the doping consists in moving
the Fermi energy into the band of peaks observed at half filling.
Thereby, the parts of the quasiparticle band which remain on the
photoemission side are essentially unaffected; the uppermost states
of this band simply cross the Fermi level and reappear as the
lowermost states of the inverse photoemission spectrum.
We have discussed the problems in directly determining the Fermi
surface from the momentum distribution function and made a situation
where they are largely avoided; then we found clear signatures of
a Fermi surface which takes the form of small hole pockets.
Both the high degree of continuity of the ground-state wave function
with decreasing $J/t$ and the identical scaling with $t/J$ of the
quasiparticle weight $Z_h$ and difference in $n({\bf k})$ between
neighboring ${\bf k}$-points suggest the existence of such
a Fermi surface also in the physical regime of parameters.
We have also found the spin-bag operators which describe the holes
dressed by the antiferromagnetic spin fluctuations, whereby elementary
excitations of the system can be described in terms of weakly-interacting
spin-1/2 Fermionic quasiparticles, corresponding to the doped holes.
We have discussed that adopting this rather conventional Fermi-liquid
scenario would explain many experimental results for high-temperature
superconductors in a very simple and natural way.

If one adopts the above Fermi-liquid picture, it seems natural to
distinguish between two types of spin excitations: the first is the
particle-hole excitation of the `hole liquid', which should resemble
that of a Fermi liquid with a Fermi-surface volume corresponding to
the number of doped holes.  In addition, there are the excitations
of the `spin background', which may, with some modifications, resemble
the spin-wave collective mode of the Heisenberg antiferromagnet.
The former part of the excitations can quantitatively explain the
Pauli susceptibility at low temperatures and the latter can explain
the dynamical spin correlations at momentum transfer $(\pi,\pi)$
observed in neutron scattering experiments.  The charge excitation
spectrum, on the other hand, resolves the internal structure of the
spin bags and observes the bare-hole excitation within the bags,
leading also to relevant physics in cuprate materials.   In the
doping region over $\sim$$30$\%, the spin and charge excitation
spectra as well as the single-particle excitation spectrum become
consistent with those of noninteracting systems with a large electronic
Fermi surface.  Details have been discussed in Refs.~[23,24].

\bigskip\noindent
{\bf Acknowledgements}
\bigskip\noindent
It is a pleasure for us to acknowledge numerous discussions with
Professor S. Maekawa.
R. E. acknowledges financial support by the Japan Society for
Promotion of Science.
This work was funded in part by Priority-Areas Grants from the
Ministry of Education, Science, and Culture of Japan.
Computations were partly carried out at the Institute for Molecular
Science, Okazaki.

\bigskip\noindent
{\bf References}
\bigskip
\frenchspacing
\item{[ 1]} R. Eder, Y. Ohta, and T. Shimozato, Phys. Rev. B {\bf 50},
            3350 (1994).
\item{[ 2]} R. Eder and Y. Ohta, Phys. Rev. B {\bf 50}, 10043 (1994).
            (also: cond-mat SISSA preprint 9406098).
\item{[ 3]} R. Eder and Y. Ohta, Phys. Rev. B, submitted.
            (also: cond-mat SISSA preprint 9407097).
\item{[ 4]} S. A. Trugman, Phys. Rev. Lett. {\bf 65}, 500 (1990).
\item{[ 5]} E. Dagotto, A. Nazarenko, and M. Boninsegni,
            Phys. Rev. Lett. {\bf 73}, 728 (1994).
\item{[ 6]} P. Aebi, J. Osterwalder, P. Schwaller, L. Schlapbach,
            M. Shimoda, T. Mochiku, K. Kadowaki, Phys. Rev. Lett. {\bf 72},
            2757 (1994).
\item{[ 7]} B. O. Wells, Z.-X. Shen, A. Matsuura, D. M. King, M. A. Kastner,
            M. Greven, and R. J. Birgeneau, preprint.
\item{[ 8]} R. Eder and P. Wr\'obel, Phys. Rev. B {\bf 47}, 6010 (1993).
\item{[ 9]} R. Eder, Phys. Rev. B {\bf 48}, 13151 (1993).
\item{[10]} R. Eder and Y. Ohta, Phys. Rev. Lett. {\bf 72}, 2816 (1994).
\item{[11]} A. Moreo and D. Duffy, preprint.
\item{[12]} W. Stephan and P. Horsch, Phys. Rev. Lett. {\bf 66}, 2258 (1991).
\item{[13]} E. Dagotto and J. R. Schrieffer, Phys. Rev. B {\bf 43},
            8705 (1990).
\item{[14]} P. Nozieres, {\it Interacting Fermi Systems} (Benjamin,
            New York, 1969).
\item{[15]} D. Poilblanc and E. Dagotto, Phys. Rev. B {\bf 42}, 4861 (1990).
\item{[16]} H. E. Castillo and C. A. Balseiro, Phys. Rev. Lett. {\bf 68},
            121 (1992).
\item{[17]} R. J. Gooding, K. J. E. Vos, and P. W. Leung, Phys. Rev. B
            {\bf 49}, 4119 (1994).
\item{[18]} A. R. Kampf and J. R. Schrieffer, Phys. Rev. B {\bf 42},
            7967 (1990).
\item{[19]} D. S. Dessau, Z.-X. Shen, D. M. King, D. S. Marshall,
            L. W. Lombardo, P. H. Dickinson, A. G. Loeser, J. Dicarlo,
            C.-H. Park, A. Kapitulnik, and W. E. Spicer, Phys. Rev. Lett.
            {\bf 71}, 2781 (1993).
\item{[20]} K. Gofron, J. C. Campuzano, H. Ding, C. Gu, R. Liu,
            B. Dabrowski, B. W. Veal, W. Cramer, and G. Jennings,
            J. Phys. Chem. Solids {\bf 54}, 1193 (1993);
            A. A. Abrikosov, J. C. Campuzano, and K. Gofron,
            Physica C {\bf 214}, 73 (1993).
\item{[21]} J. Ma, C. Quitmann, R. J. Kelley, P. Alm\'eras, H. Berger,
            G. Margaritondo, and M. Onellion, preprint.
\item{[22]} T. Nishikawa, Y. Yasui, and M. Sato, J. Phys. Soc. Jpn.,
            {\bf 63}, 3218 (1994).
\item{[23]} R. Eder and Y. Ohta, Phys. Rev. B, submitted.
            (also: cond-mat SISSA preprint 9408057).
\item{[24]} R. Eder and Y. Ohta, Phys. Rev. B, submitted.
            (also: cond-mat SISSA preprint 9409102).
\vfill
\eject
\bye